\documentclass{emulateapj}
\usepackage{natbib}
\newcommand {\aplt} {\ {\raise-.5ex\hbox{$\buildrel<\over\sim$}}\ } 

\def\sun{\hbox{$\odot$}}

\slugcomment{Accepted in ApJ Letters}
\shorttitle{GALEX and Pan-STARRS1 Discovery of SN IIP 2010aq}
\shortauthors{Gezari et al.}

\begin{document}

\title{\textsl{GALEX} and Pan-STARRS1 Discovery of SN IIP 2010\MakeLowercase{aq}: \\  The First Few Days After Shock Breakout in a Red Supergiant Star}

\author{S. Gezari,\altaffilmark{1,2} A. Rest,\altaffilmark{3} M. E. Huber,\altaffilmark{1} G. Narayan,\altaffilmark{3} K. Forster,\altaffilmark{4} J. D. Neill,\altaffilmark{4} D. C. Martin,\altaffilmark{4} S. Valenti,\altaffilmark{5} S. J. Smartt,\altaffilmark{5} R. Chornock,\altaffilmark{6} E. Berger,\altaffilmark{6} A. M. Soderberg,\altaffilmark{6} S. Mattila,\altaffilmark{7,8} E. Kankare,\altaffilmark{9,7} W. S. Burgett,\altaffilmark{10} K. C. Chambers,\altaffilmark{10} T. Dombeck,\altaffilmark{10} T. Grav,\altaffilmark{1} J. N. Heasley,\altaffilmark{10} K. W. Hodapp,\altaffilmark{10} R. Jedicke,\altaffilmark{10} N. Kaiser,\altaffilmark{10} R. Kudritzki,\altaffilmark{10} G. Luppino,\altaffilmark{10} R. H. Lupton,\altaffilmark{11} E. A. Magnier,\altaffilmark{10} D. G. Monet,\altaffilmark{12} J. S. Morgan,\altaffilmark{10} P. M. Onaka,\altaffilmark{10} P. A. Price,\altaffilmark{10} P. H. Rhoads,\altaffilmark{10} W. A. Siegmund,\altaffilmark{10} C. W. Stubbs,\altaffilmark{3} J. L. Tonry,\altaffilmark{10} R. J. Wainscoat,\altaffilmark{10} M. F. Waterson,\altaffilmark{10} and C. G. Wynn-Williams\altaffilmark{10}
}
\altaffiltext{1}{Department of Physics and Astronomy,
        Johns Hopkins University,
        3400 North Charles Street,
        Baltimore, MD 21218, USA \email{suvi@pha.jhu.edu}.}

\altaffiltext{2}{Hubble Fellow.}

\altaffiltext{3}{Department of Physics, Harvard University, Cambridge, MA 20138, USA.}

\altaffiltext{4}{California Institute of Technology, 1200 East California Blvd., Pasadena, CA  91125, USA.}

\altaffiltext{5}{Astrophysics Research Centre, School of Maths and Physics, Queen's University, BT7 1NN, Belfast, UK.}

\altaffiltext{6}{Harvard-Smithsonian Center for Astrophysics, 60 Garden Street, Cambridge, MA  02138, USA.}

\altaffiltext{7}{Tuorla Observatory, Department of Physics and Astronomy, University of Turku, FI-21500, Piikki\"o, Finland.}

\altaffiltext{8}{Stockholm Observatory, Department of Astronomy, AlbaNova University Center, SE-106 91 Stockholm, Sweden.}

\altaffiltext{9}{Nordic Optical Telescope, Apartado 474, E-38700 Santa Cruz de La Palma, Spain.}

\altaffiltext{10}{Institute for Astronomy, University of Hawaii at Manoa, Honolulu, HI 96822, USA.}

\altaffiltext{11}{Department of Astrophysical Science, Princeton University, Princeton, NJ  08544, USA.}

\altaffiltext{12}{US Naval Observatory, Flagstaff Station, Flagstaff, AZ 86001, USA.}

\begin{abstract}
We present the early UV and optical light curve of Type IIP supernova (SN) 2010aq at $z=0.0862$, and compare it to analytical models for thermal emission following SN shock breakout in a red supergiant star.  SN 2010aq was discovered in joint monitoring between the \textsl{Galaxy Evolution Explorer (GALEX)} Time Domain Survey (TDS) in the NUV and the Pan-STARRS1 Medium Deep Survey (PS1 MDS) in the $g$, $r$, $i$, and $z$ bands.    The \textsl{GALEX} and Pan-STARRS1 observations detect the SN less than 1 day after shock breakout, measure a diluted blackbody temperature of $31,000 \pm 6,000$ K 1 day later, and follow the rise in the UV/optical light curve over the next 2 days caused by the expansion and cooling of the SN ejecta.  The high signal-to-noise ratio of the simultaneous UV and optical photometry allows us to fit for a progenitor star radius of $700 \pm 200 R_{\odot}$, the size of a red supergiant star.  An excess in UV emission two weeks after shock breakout compared to SNe well fitted by model atmosphere-code synthetic spectra with solar metallicity, is best explained by suppressed line blanketing due to a lower metallicity progenitor star in SN 2010aq.  Continued monitoring of PS1 MDS fields by the \textsl{GALEX} TDS will increase the sample of early UV detections of Type II SNe by an order of magnitude, and probe the diversity of SN progenitor star properties. 
\end{abstract}


\section{Introduction}
Shock breakout in a core-collapse supernova (SN) marks the first escape of radiation from the blast wave that breaks through the surface of the star and launches the SN ejecta, heating it to temperatures of $10^{5}-10^{6}$ K \citep{Colgate1974, Falk1978, Klein1978, Matzner1999}.  The duration of this radiative precursor in the UV/X-rays is smeared out by light-travel time effects to $t_{lt}=R_{\star}/c \sim 2$ s $(R_{\star}/R_{\odot})$, where $R_{\star}$ is the pre-SN radius of the progenitor star.  In the presence of a dense wind, or if the shock breakout is aspherical or very weak, the observed duration can be much longer \citep{Campana2006, Chevalier2008, Couch2009, DessartLivne2010}.

Following shock breakout, the early SN light curve is powered by radiation from the expanding and cooling SN ejecta.  In hydrogen-rich Type II-plateau SNe (SNe IIP), a receding H recombination wave prolongs the optical brightness of the SN until radioactive heating plays a dominant role.  The luminosity and temperature evolution of the early thermal expansion phase can be used as a diagnostic for the radius of the progenitor star, as well as the explosion energy to ejecta mass ratio \citep{Waxman2007}.  This is an important tool, since the direct detection of SN progenitors is incredibly difficult, and has only been possible for a small number of nearby SNe with pre-explosion high resolution imaging \citep{Smartt2009}.    

Recent observational advances have enabled the detection of early UV emission from core-collapse SNe from hours to a couple of days after explosion.  The \textsl{Swift} satellite serendipitously detected luminous X-ray bursts with X-Ray Telescope (XRT) from Type Ib/c SNe 2006aj and 2008D, which were modeled by shock breakout in the dense wind surrounding a Wolf-Rayet (WR) star \citep{Campana2006, Waxman2007, Soderberg2008}, although alternative scenarios were also proposed that invoked the presence of relativistic jets \citep{Ghisellini2007, Mazzali2008}.  The UV/optical counterparts detected in co-aligned \textsl{Swift} UVOT observations were well fitted by analytical models for emission from the expanding, cooling envelope of a progenitor with the radius of a WR star \citep{Campana2006, Soderberg2008, Rabinak2010}. 

The wide-field UV survey capabilities of the \textsl{Galaxy Evolution Explorer (GALEX)} satellite serendipitously detected the onset within 7 hr of shock breakout of a UV-bright plateau lasting $2$ d before fading away from Type IIP SN SNLS-04D2dc at $z=0.185$, two weeks before its discovery in the optical Supernova Legacy Survey (SNLS) \citep{Gezari2008, Schawinski2008}.  While the data were modeled both numerically and analytically with thermal emission following shock breakout in an RSG progenitor with $R_{\star}$ ranging from $800$ to $1000 R_{\odot}$ and a variety of stellar envelope density structures \citep{Gezari2008, Schawinski2008, Tominaga2009, Rabinak2010}, the low signal-to-noise ratio of the detections (S/N$ < 3$) and the lack of simultaneous optical data preclude one from fitting for the parameters of the RSG progenitor star.    

In this Letter, we present SN IIP 2010aq at $z=0.0862$ discovered in joint monitoring observations in the UV by \textsl{GALEX} Time Domain Survey (TDS) and in the optical by the Pan-STARRS1 Medium Deep Survey (PS1 MDS).  The UV and optical observations discover the SN less than 20.5 hr after shock breakout in the SN rest frame, and the high S/N of the detections enable us to use model fits to the early light curve to place meaningful constraints on the radius of the RSG progenitor star for the first time.    

\section{Observations}

From UT 2010 February 4 - March 22 (UT dates are used throughout this Letter) the \textsl{GALEX} TDS monitored the 7 deg$^{2}$ PS1 MDS field MD04 in the NUV with 7 pointings, each 1 deg$^{2}$, with a cadence of 2 days and with a limiting magnitude of $m_{\rm AB} \sim 23$.  Gaps in the regular monitoring occurred when the NUV detector was shut down for engineering tests.  The \textsl{GALEX} science images were run through the difference imaging pipeline {\tt photpipe} \citep{Rest2005}, and difference image detections were filtered through a series of selection criteria to remove false detections due to image-subtraction artifacts.  The remaining detections were associated between epochs, and then cross-matched with the PS1 SN alerts from the MDS with a matching radius of 5$\arcsec$.  

\begin{figure}
\plotone{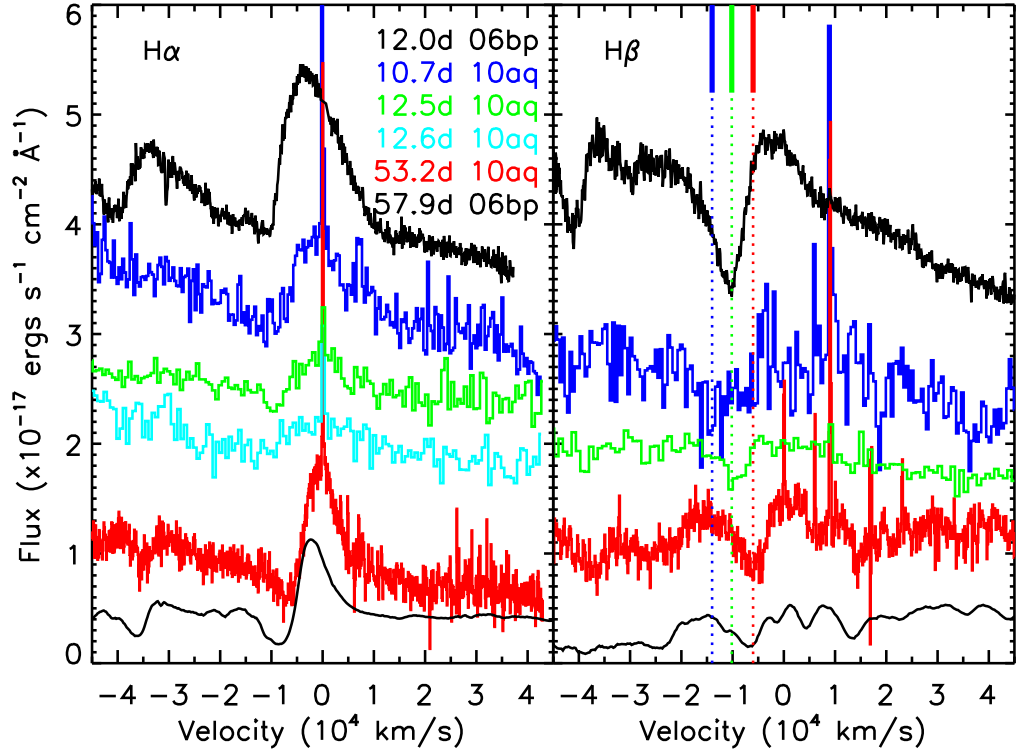}
\caption{Strong H$\alpha$ and H$\beta$ P-Cygni features in SN 2010aq, with the velocity of the H$\beta$ absorption minimum (which shifts from 14,000 km s$^{-1}$ to 6,000 km s$^{-1}$ between $\phi = 10.7$ and $53.2$ d) marked with a dotted line.  Also shown is the $\phi = 12.0$ d and $\phi =57.9$ d spectrum for SN 2006bp from \citet{Dessart2008} and \citet{Quimby2007} which have been multiplied by a factor of 0.013 and 0.002, respectively, and have an H$\beta$ absorption minimum at $10,160$ km s$^{-1}$ and $7,000$ km s$^{-1}$, respectively.  The spectra have been binned in wavelength when necessary to improve the S/N, and offset in flux for clarity.
\label{fig:vel}
}
\end{figure}

\textsl{GALEX} TDS detected a variable object on 2010 February 12 coincident with a PS1 SN candidate discovered at R.A.$ = 10^{\rm h} 02^{\rm m} 09\fs741$, decl.$ = +01\degr 14\arcmin 00\farcs94$ (J2000) in daily imaging cycling through the $g$, $r$, $i$, and $z$ filters starting on 2010 February 12.  The SN was confirmed with spectroscopy with the 4.2m William Herschel Telescope (WHT) with ISIS on 2010 February 23.08 to be a Type IIP SN at $z=0.0862$ from its strong Balmer P-Cygni features \citep[SN 2010aq,][]{Valenti2010}.  We obtained additional epochs of spectroscopy with the Nordic Optical Telescope (NOT) with ALFOSC on 2010 February 24.98, the 6.5 m Magellan Telescope with IMACS on 2010 February 25.17, and with the 6.5m MMT with Hectospec on 2010 April 10.20. In Figure \ref{fig:vel} we show the H$\alpha$ and H$\beta$ features in our spectra, and the velocity drift of the absorption minimum of H$\beta$ with time from the recession of the photosphere in mass.

In order to avoid systematic errors from residuals in the difference images, the \textsl{GALEX} photometry is measured on the science images with a 6$\arcsec$ radius aperture centered on the host galaxy, and the flux of the SN is determined by subtracting the flux of the host galaxy measured from pre-SN epochs ($NUV=22.0 \pm 0.1$ mag).  The 1$\sigma$ error is measured empirically from the dispersion in magnitudes measured for sources detected in all epochs.  The offset between the SN and the host galaxy centroid is $1\farcs5$ (1 pixel), for which we estimate an upper limit to systematic errors due to centering of the aperture $<$0.1 mag.  The \textsl{GALEX} photometry is in the AB system, has been aperture corrected \citep[$\Delta m=-0.23$,][]{Morrissey2007}, and is corrected for Galactic extinction with $E(B-V)=0.024$ mag.

The PS1 MDS images were reduced with the PS1 Image Processing Pipeline \citep{Magnier2006} including detrending, calibration, warping, and stacking.  Difference imaging and transient alerting were performed using the {\tt photpipe} pipeline on the stacked PS1 images \citep{Rest2005}. The SN photometry is measured by {\tt photpipe} using forced point-spread function (PSF) photometry on the difference images at the position of the SN, and the zeropoint is measured from comparison with field stars in the SDSS catalog.  The PS1 photometry is in the AB system, and is corrected for Galactic extinction.

Table \ref{tab1} lists the \textsl{GALEX} and PS1 SN photometry, and Figure \ref{fig:lc} shows the UV and optical SN light curve.  We define the phase of the SN ($\phi$) to be days in the SN rest frame since the time of shock breakout ($t_{\rm SBO}$).  Our observed constraint on $t_{\rm SBO}$ is the time between the last non-detection with \textsl{GALEX} in the NUV on MJD $55237.54$, and the first detection with PS1 in the $z$-band on MJD $55238.47$, which is a window of 20.5 h in the SN rest frame.  We further constrain $t_{\rm SBO}$ using fits of shock breakout models to the early UV/optical light curve in \S \ref{sec:models} to be $t_{\rm SBO}=$ MJD $55238.42 \pm 0.02$.  

\section{Host Galaxy}

We measure the host galaxy redshift of $z=0.0862$ from nebular emission lines that are superimposed on the SN spectrum, and adopt a distance modulus, DM=38.0. The narrow Balmer lines have a Galactic-extinction corrected H$\alpha$/H$\beta$ ratio of $3.28 \pm 0.15$, corresponding to an internal extinction of $E(B-V$)=0.12 $\pm$ 0.04 mag \citep{Calzetti2001}.  We use the semiempirical relation between metallicity and the measured nebular emission line ratios, [N II]$\lambda 6583$/H$\alpha < 0.1$, [O III]$\lambda 5007/$H$\beta = 3.6 \pm 0.2$, and $\log(R_{23} = $([O II]$\lambda 3727 + $[O III]$\lambda 5007, 4959$)/H$\beta) = 0.89 \pm 0.05$, to measure a low oxygen abundance of $12+$log(O/H)$ = 8.2-8.5$ \citep{McGaugh1991, vanZee1998, Pettini2004}, which is consistent with the low metallicity expected from the luminosity-metallicity relation from \citet{Tremonti2004} given the dwarf host galaxy of SN 2010aq with an internal extinction and $K$-corrected rest-frame absolute magnitude of $M_{B}=-17.3$ mag.

\section{Comparison with Analytical Models} \label{sec:models}

We compare the early ($\phi < 3$ d) UV/optical light curve to analytical models for shock breakout and its immediate aftermath in a RSG star, the progenitors of Type IIP SNe.  The analytical model presented by \citet{Nakar2010} (NS10), follows the photon-gas coupling to determine the location of the photosphere, and takes into account photon diffusion from inner layers below the photosphere.  The analytical relations presented by \citet{Rabinak2010} (RW10) neglect the effects of photon diffusion, but do include opacity from electron scattering and its effect on the emitted spectrum.  Although the pre-SN stellar structure and detailed physics are better treated in numerical radiation hydrodynamics models, these models are computing-time intensive, and analytical models are more practical to use for fitting for the parameters of the progenitor star.  We rely on the analytical scaling relations from NS10 to constrain $R_{500}$ and $E_{51}/M_{15}$, where $R_{500}$ is $R_{\star}/500 R_{\odot}$, $M_{15}=M/15 M_{\odot}$, and $E_{51}= E/10^{51}$ ergs.  Here $E$ is the energy of the explosion, and $M$ refers to the mass of the ejecta, and does not include the 1.44 $M_{\odot}$ from the neutron star core or mass lost from winds during the lifetime of the star.  

In the spherical expansion phase (when $R_{ph} \approx v_{ph}t$), the NUV light curve rises to a maximum when the ejecta expands and cools until the peak energy of the emission is $3kT = h\nu_{NUV}(1+z)$, or when $T = 1.94$ eV.  Using $T(\phi)$ for a RSG from NS10, this translates to $\phi_{peak}=2.2$ d $M_{15}^{-0.23} R_{500}^{0.68} E_{51}^{0.20}$.  For a blue supergiant star (BSG) with $R =20-70 R_{\odot}$, $\phi_{peak}$ in the NUV occurs much earlier ($\phi_{peak} < 0.6$ d), which is incompatible with the observed light curve of SN 2010aq.  If we use our observed limits on $t_{\rm SBO}$= MJD $55238.0 \pm 0.5$, and conservatively estimate $\phi_{peak}$ to be within 1 day of the second \textsl{GALEX} detection on MJD 55241.5, then in the SN rest-frame $\phi_{peak} = 3 \pm 1$ d, and this results in $M_{15}^{-0.23}R_{500}^{0.68}E_{51}^{0.20} = 1.4 \pm 0.5$.  

The first detection in the NUV ($\phi_{1}$) on MJD 55239.30, as well as on MJD 55239.46 in the $g$ and $r$ bands, places a tight constraint on $T(\phi_{1}$).  In the first couple of days after shock breakout, the spectrum of the emission is well approximated by a diluted blackbody \citep{Gezari2008, Tominaga2009}.  Assuming that the $g$ and $r$ data points are $\sim 0.2$ mag fainter on MJD 55239.30, the UV/optical photometry is fitted with $E(B-V)=0.11 \pm 0.01$ mag and $T_{bb}=2.7 \pm 0.5$ eV, in excellent agreement with $E(B-V)$ derived from the Balmer decrement.  Note that the observed diluted blackbody temperature is hotter than the photospheric temperature by a factor of $\approx 1.2$ as a result of electron scattering (RW10).

In Figure \ref{fig:lc_early} we plot the NS10 model for an RSG with $M_{15} = R_{500} = E_{51} = 1$, fitted for $E(B-V)=0.03$ mag and $t_{sbo} = $MJD 55238.42 $\pm 0.02$, with an offset of $-1.5$ mag and a K correction to transform the NS10 models from the rest-frame to the observed frame of SN 2010aq.  The NS10 model is in good agreement with the data until $\phi=3$ d. The tight constraint on $t_{\rm SBO}$ is from the fit to the earliest detection in the $z$-band on $\phi = 0.03 \pm 0.02$ d, and does not include the much larger uncertainties at very early phases ($\phi $\aplt$ 1.5$ d for a RSG) introduced by the simplistic approximations in the analytical models (RW10).  
While the NS10 $M_{15} = R_{500} = E_{51} = 1$ model is fitted to the data with a small amount of reddening, if we require that $E(B-V)=0.12 \pm 0.04$ mag, then one gets the same fit at $\phi_{1}$ to a model with a factor of $1.4^{+0.4}_{-0.2}$ times higher temperature, and thus with $M_{15}^{-0.13}R_{500}^{0.38}E_{51}^{0.11}=1.4^{+0.4}_{-0.2}$.    Given the offset in luminosity required to fit the model, and $L(\phi)$ for an RSG from NS10, one gets an additional constraint of $M_{15}^{-0.87}R_{500}E_{51}^{0.96} \sim 4$. 

When combining all of the constraints above, an extinction of $E(B-V)<0.12$ mag is required for the allowed parameters from $\phi_{peak}$ to converge with those from $T(\phi_{1})$.  If we assume the best fit value for the reddening from $T(\phi_{1})$ of $E(B-V)=0.11 \pm 0.01$ mag, which is in agreement with this upper limit, then for different values of $E_{51}$, one then gets the following solutions, ($E_{51}, M_{15}, R_{500}$) = (0.5, $0.09-0.15$, $1.0-1.5$), (1.0, $0.23-0.38$, $1.1-1.8$), and (1.5, $0.37-0.62$, $1.1-1.8$), thus placing the strongest constraint on the radius of the RSG progenitor, $R_{500}=1.4 \pm 0.4$.  For this value of $R_{\star}$, stellar evolutionary models of core-collapse SN progenitor stars with solar to half-solar metallicities have $M_{15}\sim 0.45-0.90$ \citep{Eldridge2004}, and hydrodynamical simulations of SNe IIP have even larger model ejecta masses ($M_{15} > 1$) \citep{Utrobin2008}, thus favoring the solutions with $E_{51}=$1.5.

In Figure \ref{fig:lc_early}, we also show a $K$-corrected analytical model with a reddening of $E(B-V) = 0.03$ mag from RW10 with parameters within the limits of the allowed solutions ($E_{51}=1.44, M_{15}=0.59,$ and $R_{500}=1.73$) that has a luminosity and UV/optical colors that are in good agreement with the data at $\phi_{1}$, but with a slower temperature evolution that is in slightly worse agreement with the data at $\phi = 3$ d.  If we again assume $E(B-V)$ = 0.11 mag, then the model luminosity increases by a factor of 1.8, resulting in a similar increase in $E_{51}/M_{15}$ for values of $R_{500} < 1.8$.

\section{Comparison with Type IIP SN\MakeLowercase{e} 2005\MakeLowercase{cs} and 2006\MakeLowercase{bp}}

\begin{figure}
\plotone{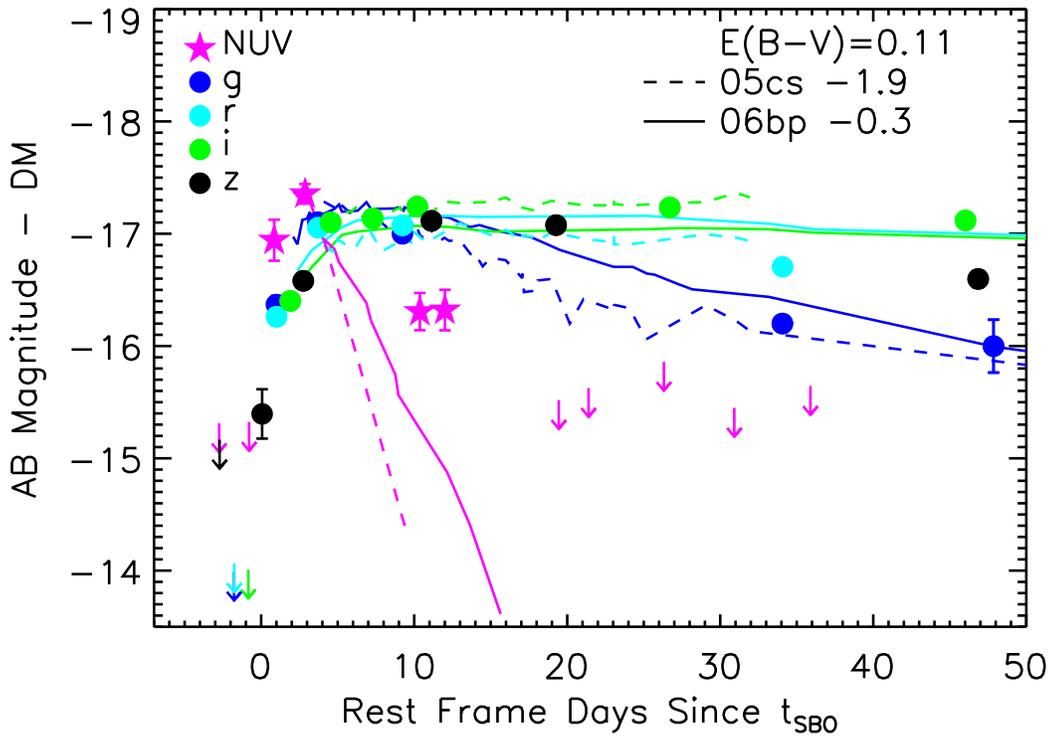}
\caption{\textsl{GALEX} NUV and PS1 $g, r, i, $ and $z$ light curve of SN 2010aq.  The flux of the host galaxy has been removed, and the magnitudes are in the AB system, corrected for Galactic extinction, and subtracted by the distance modulus (DM=38.0).  Solid and dashed lines show the light curves of well-studied Type IIP SNe 2005cs and 2006bp from \citet{Dessart2008}, and $K$-corrected to the observed frame of SN 2010aq.  SN 2005cs was reddened by 0.02 mag and SN 2006bp was dereddened by 0.24 mag to remain with a net internal reddening of $E(B-V)=0.11 \pm 0.05$ mag and offset by $-1.9$ and $-0.3$ mag, respectively.
\label{fig:lc}
}
\end{figure}

\begin{figure}
\plotone{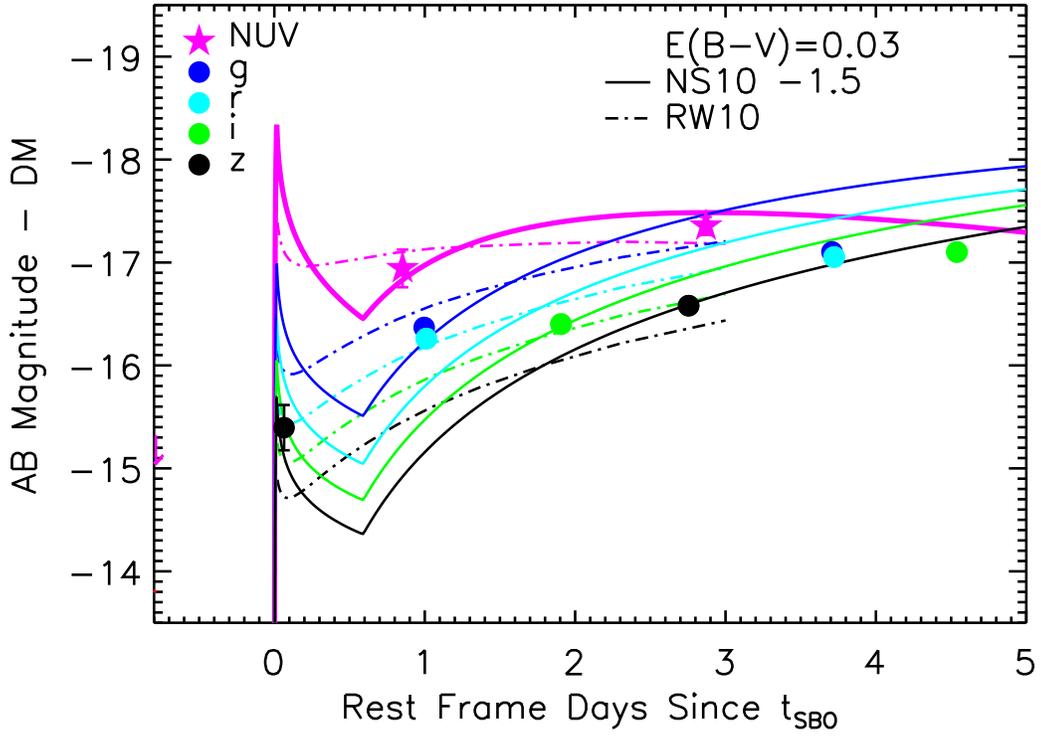}
\caption{Zoom in of the early UV/optical light curve of SN 2010aq in comparison to the best fitting analytical model from \citet{Nakar2010} (NS10) with  $E_{51} = 1$, $M_{15} = 1$, and $R_{500}=1$ (thick solid line), and the analytical model from \citet{Rabinak2010} (RW10) with $E_{51}=1.44$, $M_{15}=0.59$, and $R_{500}= 1.73$ (dashed-dotted line).  The models have been $K$-corrected into the observed frame of SN 2010aq and applied with a reddening of $E(B-V)=0.03$ mag.  The NS10 model has been offset by $-1.5$ mag.
\label{fig:lc_early}
}
\end{figure}

We compare the late-time ($\phi > 3$~d) UV/optical light curve of SN 2010aq with two nearby Type IIP SNe well-studied in the UV and optical with the \textsl{Swift} satellite and ground-based photometry and spectroscopy, SN 2005cs (DM=29.7) and SN 2006bp (DM=31.2) \citep[see summary in][]{Dessart2008}.  In order to compare the light curves to SN 2010aq, we use SN IIP template spectra to measure the K correction required to transform the observations into the \textsl{GALEX} and PS1 filters in the observed frame of SN 2010aq, and apply a net internal reddening of $E(B-V)=0.11$ mag, where we use the \citet{Cardelli1988} law, and the internal reddening estimated for SN 2005cs (2006bp) from synthetic spectral fits to optical spectra and UV photometry of $E(B-V)=0.04$ ($0.40$) mag, with an accuracy of $\pm 0.05$ mag from \citet{Dessart2008}.
We set $\phi=0$ for SN 2005cs (2006bp) to the x-intercept of the expanding photosphere from \citet{Dessart2008} of MJD 53547.3 (MJD 53832.9), and after subtracting the distance modulus, shift the light curves by $-1.9$ mag ($-0.3$ mag).  The optical light curves are in good agreement between the SNe in Figure \ref{fig:lc}, and the similar evolution of the photospheric velocity with time of SNe 2010aq and 2006bp (see comparison in Figure \ref{fig:vel}), suggests they have a similar $E/M$, since $v_{ph} \propto \sqrt{E/M}$.  

Figure \ref{fig:lc} shows a large NUV excess of SN 2010aq compared to SNe IIP 2005cs and 2006bp at $\phi =11-13$ d.  
Given that the NUV excess is more prominent at $\phi=11-13$ d than at $\phi=3$ d, and that the $\phi=10.7$ d spectrum does not show signs of circumstellar interaction, it is most likely that the NUV excess is associated with the effects of line blanketing by metal ions shortward of 3000 \AA, which increases with time when the photospheric temperature of the SN is cool enough for Fe II to be present \citep{Dessart2005}.  Indeed, if the metallicity of the solar-metallicity non-LTE atmosphere-code synthetic spectra fitted to SNe 2005cs and 2006bp from \citet{Dessart2008} was decreased, there would be a significant increase in the emitted NUV flux \citep[see Figure 9,][]{Dessart2005}.  However, one must be cautious with this interpretation, since the UV spectral range is difficult to model. Even in the case of SN 1987A, where the metallicity, BSG progenitor, and explosion energy of the SN are well constrained, the most sophisticated time-dependent non-LTE radiative transfer models, while in excellent agreement with the optical observations (at the level of $10$\%), are in much worse agreement in the UV (only to the $30-50$\% level) \citep{Dessart2010}.  However, given the low metallicity of the dwarf host galaxy of SN 2010aq, a subsolar metallicity progenitor star is the most natural explanation for the UV excess at late times. 

\section{Conclusions}

Although the duration of the radiative precursor from shock breakout is too short to be resolved by rolling SN surveys, the thermal expansion phase in the days following can be studied in detail.  In this Letter, we have shown that with daily cadence UV and optical imaging we can use analytical models to fit for the radius of the pre-SN progenitor star, and even use the evolution of the UV light curve to directly probe its metallicity.  UV observations are also critical for constraining the time of shock breakout in distant SNe ($t_{peak}-t_{\rm SBO}$ is only 2 days in the NUV compared to 2 weeks in the $r$ band), and can improve the accuracy of using Type II-P SNe as distance indicators \citep{Schmidt1992, Dessart2008, Kasen2009}. 

Based on the detection efficiency of \textsl{GALEX} for detecting UV emission from Type II SNe measured from serendipitous observations of SNLS SN candidates, \citet{Gezari2008} predict a detection rate for \textsl{GALEX} TDS, which monitors one PS1 MD field per month, of $\approx 1$ month$^{-1}$.  Given the actual detections of two SNe IIP in effectively 2 months of operations of \textsl{GALEX} TDS in coordination with PS1, luminous SN 2009kf at $z=0.182$ \citep{Botticella2010} and SN 2010aq, our detection rate appears to be right on the mark.  Continued  \textsl{GALEX} TDS and PS1 MDS monitoring will yield an exciting data set to probe the diversity of progenitors of core-collapse SNe, and approach a better understanding of the mapping from stellar evolution to stellar death.
 
\acknowledgements
S.~G. thanks I.~Rabinak and E.~Nakar for kindly providing their models in the \textsl{GALEX} and PS1 filters, L.~Dessart for helpful discussions, and the anonymous referee for useful comments.  S.~G. was supported by NASA through Hubble Fellowship grant HST-HF-01219.01-A awarded by the Space Telescope Science Institute, which is operated by AURA, Inc., for NASA, under contract NAS 5-26555.  The PS1 Surveys have been made possible through the combinations of the Institute for Astronomy at the University of Hawaii, The Pan-STARRS Project Office, the Max-Planck Society and its participating institutes, the Max Planck Institute for Astronomy, Heidelberg, and the Max Planck Institute for Extraterrestrial Physics, Garching, The Johns Hopkins University, the University of Durham, the University of Edinburgh, the Queen's University of Belfast, the Harvard-Smithsonian Center for Astrophysics, the Las Cumbres Observatory Global Network, and the National Central University of Taiwan.  We gratefully acknowledge NASA's support for construction, operation, and science analysis for the \textsl{GALEX} mission, developed in cooperation with CNES of France and the Korean MOST. 


\begin{thebibliography}{}

\bibitem[Botticella et al.(2010)]{Botticella2010}
Botticella, M. T., et al. 2010, ArXiv e-prints, 1001.5427

\bibitem[Calzetti(2001)]{Calzetti2001}
Calzetti, D. 2001, PASP, 113, 1449

\bibitem[Campana et al.(2006)]{Campana2006}
Campana, S., et al. 2006, Nature, 442, 1008

\bibitem[Cardelli et al.(1988)]{Cardelli1988}
Cardelli, J. A., Clayton, G. C., \& Mathis, J. S. 1988, ApJ, 329,
L33

\bibitem[Chevalier \& Fransson(2008)]{Chevalier2008}
Chevalier, R. A. \& Fransson, C. 2008, ApJ, 683, L135

\bibitem[Colgate(1974)]{Colgate1974}
Colgate, S. A. 1974, ApJ, 187, 333

\bibitem[Couch et al.(2009)]{Couch2009}
Couch, S. M., Wheeler, J. C., \& Milosavljevi´c, M. 2009, ApJ, 696,
953

\bibitem[Dessart et al.(2008)]{Dessart2008}
Dessart, L., et al. 2008, ApJ, 675, 644

\bibitem[Dessart \& Hillier(2005)]{Dessart2005}
Dessart, L., \& Hillier, D. J. 2005, A\&A, 437, 667

\bibitem[Dessart \& Hillier(2010)]{Dessart2010}
Dessart, L., \& Hillier, D. J. 2010, MNRAS, 405, 2141

\bibitem[Dessart et al.(2010)]{DessartLivne2010}
Dessart, L., Livne, El, \& Waldman, R. 2010, MNRAS, 405, 2113

\bibitem[Eldridge \& Tout(2004)]{Eldridge2004}
Eldridge, J. J., \& Tout, C. A. 2004, MNRAS, 353, 87

\bibitem[Falk(1978)]{Falk1978}
Falk, S. W. 1978, ApJ, 225, L133

\bibitem[Gezari et al.(2008)]{Gezari2008}
Gezari, S., et al. 2008, ApJ, 683, L131

\bibitem[Ghisellini et al.(2007)]{Ghisellini2007}
Ghisellini, G., Ghirlanda, G., \& Tavecchio, F. 2007, MNRAS, 382, L77

\bibitem[Kasen \& Woosley(2009)]{Kasen2009}
Kasen, D., \& Woosley, S. E. 2009, ApJ, 703, 2205

\bibitem[Klein \& Chevalier(1978)]{Klein1978}
Klein, R. I., \& Chevalier, R. A. 1978, ApJ, 223, L109

\bibitem[Magnier(2006)]{Magnier2006}
Magnier, E. 2006, in The Advanced Maui Optical and Space
Surveillance Technologies Conference

\bibitem[Matzner \& McKee(1999)]{Matzner1999}
Matzner, C. D., \& McKee, C. F. 1999, ApJ, 510, 379

\bibitem[Mazzali et al.(2008)]{Mazzali2008}
Mazzali, P., et al. 2008, Science, 321, 1185

\bibitem[McGaugh(1991)]{McGaugh1991}
McGaugh, S. S. 1991, ApJ, 380, 140

\bibitem[Morrissey et al.(2007)]{Morrissey2007}
Morrissey, P., et al. 2007, ApJS, 173, 682

\bibitem[Nakar \& Sari(2010)]{Nakar2010}
Nakar, E., \& Sari, R. 2010, ArXiv e-prints, 1004.2496



\bibitem[Pettini \& Pagel(2004)]{Pettini2004}
Pettini, M., \& Pagel, B. E. J. 2004, MNRAS, 348, L59

\bibitem[Quimby et al.(2007)]{Quimby2007}
Quimby, R. M., Wheeler, J. C., H¨oflich, P., Akerlof, C. W., Brown,
P. J., \& Rykoff, E. S. 2007, ApJ, 666, 1093

\bibitem[Rabinak \& Waxman(2010)]{Rabinak2010}
Rabinak, I., \& Waxman, E. 2010, ArXiv e-prints, 1002.3414

\bibitem[Rest et al.(2005)]{Rest2005}
Rest, A., et al. 2005, ApJ, 634, 1103

\bibitem[Schawinski et al.(2008)]{Schawinski2008}
Schawinski, K., et al. 2008, Science, 321, 223

\bibitem[Schmidt et al.(1992)]{Schmidt1992}
Schmidt, B. P., Kirshner, R. P., \& Eastman, R. G. 1992, ApJ, 395,
366

\bibitem[Smartt et al.(2009)]{Smartt2009}
Smartt, S. J., Eldridge, J. J., Crockett, R. M., \& Maund, J. R.
2009, MNRAS, 395, 1409

\bibitem[Soderberg et al.(2008)]{Soderberg2008}
Soderberg, A. M., et al. 2008, Nature, 453, 469

\bibitem[Tominaga et al.(2009)]{Tominaga2009}
Tominaga, N., Blinnikov, S., Baklanov, P., Morokuma, T., Nomoto,
K., \& Suzuki, T. 2009, ApJ, 705, L10

\bibitem[Tremonti et al.(2004)]{Tremonti2004}
Tremonti, C. A., et al. 2004, ApJ, 613, 898

\bibitem[Utrobin \& Chugai(2008)]{Utrobin2008}
Utrobin, V. P. \& Chugai, N. N. 2008, A\&A, 491, 507

\bibitem[Valenti et al.(2010)]{Valenti2010}
Valenti, S., et al. 2010, Central Bureau Electronic Telegrams, 2214,
1

\bibitem[van Zee et al.(1998)]{vanZee1998}
van Zee, L., Salzer, J. J., Haynes, M. P., O’Donoghue, A. A., \&
Balonek, T. J. 1998, AJ, 116, 2805

\bibitem[Waxman et al.(2007)]{Waxman2007}
Waxman, E., M´esz´aros, P., \& Campana, S. 2007, ApJ, 667, 351

\end{thebibliography}

\clearpage

\begin{deluxetable}{ccrcc}
\tablewidth{0pt}
\tablecaption{\textsl{GALEX} and PS1 Photometry\label{tab1}}
\tablehead{
\colhead{Filter} & \colhead{Date (MJD)} & \colhead{$\phi$ (d)\tablenotemark{a}} & \colhead{mag \tablenotemark{b}} & \colhead{1$\sigma$}
}
\startdata
NUV & 55235.42 &  $-$2.8 & (22.69) & \nodata \\
NUV & 55237.54 &  $-$0.8 & (22.68) & \nodata \\
NUV & 55239.32 &   0.8 & 21.06 & 0.18 \\
NUV & 55241.52 &   2.8 & 20.64 & 0.09 \\
NUV & 55249.66 &  10.3 & 21.69 & 0.17 \\
NUV & 55251.44 &  12.0 & 21.68 & 0.18 \\
NUV & 55259.52 &  19.4 & (22.48) & \nodata \\
NUV & 55261.64 &  21.4 & (22.37) & \nodata \\
NUV & 55266.98 &  26.3 & (22.14) & \nodata \\
NUV & 55271.98 &  30.9 & (22.55) & \nodata \\
NUV & 55277.39 &  35.9 & (22.36) & \nodata \\
g & 55236.48 & $-$1.8 & (24.00) & \nodata \\
g & 55239.48 &   1.0 & 21.63 & 0.08 \\
g & 55242.43 &   3.7 & 20.90 & 0.03 \\
g & 55248.43 &   9.2 & 21.00 & 0.03 \\
g & 55275.39 &  34.0 & 21.80 & 0.07 \\
g & 55290.38 &  47.8 & 22.00 & 0.24 \\
g & 55305.27 &  61.5 & 23.03 & 0.32 \\
r & 55236.49 & $-$1.79 & (23.90) & \nodata \\
r & 55239.50 &   1.0 & 21.74 & 0.07 \\
r & 55242.44 &   3.7 & 20.95 & 0.02 \\
r & 55248.44 &   9.2 & 20.93 & 0.03 \\
r & 55275.41 &  34.0 & 21.29 & 0.05 \\
r & 55305.29 &  61.6 & 21.92 & 0.11 \\
i & 55237.49 & $-$0.9 & (23.99) & \nodata \\
i & 55240.47 &   1.9 & 21.60 & 0.05 \\
i & 55243.33 &   4.5 & 20.90 & 0.03 \\
i & 55246.32 &   7.3 & 20.86 & 0.04 \\
i & 55249.48 &  10.2 & 20.76 & 0.03 \\
i & 55267.41 &  26.7 & 20.77 & 0.05 \\
i & 55288.39 &  46.0 & 20.88 & 0.05 \\
i & 55306.25 &  62.4 & 21.64 & 0.09 \\
i & 55321.27 &  76.3 & 22.42 & 0.14 \\
z & 55235.45 & $-$2.7 & (22.83) & \nodata \\
z & 55238.47 & 0.0 & 22.60 & 0.22 \\
z & 55241.39 &   2.7 & 21.42 & 0.07 \\
z & 55250.49 &  11.1 & 20.89 & 0.06 \\
z & 55259.33 &  19.2 & 20.92 & 0.05 \\
z & 55289.29 &  46.8 & 21.40 & 0.07 \\
z & 55298.34 &  55.2 & 21.53 & 0.10 \\
z & 55301.37 &  57.9 & 20.98 & 0.16 \\
z & 55304.26 &  60.6 & 21.70 & 0.08
\enddata
\tablenotetext{a}{Days since MJD 55238.4 in the SN rest-frame.}
\tablenotetext{b}{AB magnitudes corrected for a Galactic extinction of E(B-V)=0.024 and the host galaxy contribution removed.  Upper limits are given in parentheses.}
\end{deluxetable}

\end{document}